\documentclass{PoS}
\usepackage[utf8]{inputenc}
\usepackage{amsmath}
\usepackage{amssymb}
\usepackage{graphicx}
\usepackage[capitalize]{cleveref}

\graphicspath{{./graphics/}}

\title{Matter And Gravitation In Collisions of heavy ions and neutron stars: equation of state}

\ShortTitle{Matter And Gravitation In Collisions}

\author{\mainauthor{Anton~Motornenko}$^{1,2}$, Jan~Steinheimer$^{1}$, Volodymyr~Vovchenko$^{1,2}$,}
\author{\deceased{Stefan~Schramm}$^{1,2}$, and  \speaker{Horst~Stoecker}$^{1,2,3}$\\
$^1$ Institut f\"ur Theoretische Physik,
Goethe Universit\"at, D-60438 Frankfurt am Main, Germany\\
$^2$ Frankfurt Institute for Advanced Studies,
Giersch Science Center, D-60438 Frankfurt am Main, Germany\\
$^3$ GSI Helmholtzzentrum f\"ur Schwerionenforschung GmbH, D-64291 Darmstadt, Germany\\
E-mail: \email{motornenko@fias.uni-frankfurt.de}}

\abstract{
The gravitational waves emitted from a binary neutron star merger, as predicted from general relativistic magneto- hydrodynamics  calculations, are sensitive to the
appearance of quark matter and the stiffness of the equation of state of QCD matter present in the inner cores of the stars. 
This is a new messenger observable from outer space,
which does provide direct signals for the phase structure of strongly interacting QCD
matter at high baryon density and high temperature.
These astrophysically created extremes of thermodynamics do match, to within 20\%,
the values of densities and temperatures which we find in relativistic hydrodynamics
and transport theory of heavy ion collisions at the existing laboratories,
if though at quite different rapidity windows, impact
parameters and bombarding energies of the heavy nuclear systems.
We demonstrate how one unified equation of state can be constructed and used for both neutron star physics and hot QCD matter excited at laboratory facilities. The similarity in underlying QCD physics allows the gravitational wave signals from future advanced LIGO and Virgo
events to be combined with the analysis of
high multiplicity fluctuations and flow measurements in
heavy ion detectors in the lab to pin down the EoS and the phase structure of dense matter. }

\FullConference{Corfu Summer Institute 2018 "School and Workshops on Elementary Particle Physics and Gravity"\\
		(CORFU2018)\\
		31 August - 28 September, 2018\\
		Corfu, Greece}

\begin{document}

\section{\label{sec:int}Introduction}

Modern theory of strong interactions, Quantum Chromodynamics (QCD), is challenged by two actively developing divisions of experimental physics. On one side the accelerator driven facilities are spending efforts in measuring properties of systems created in relativistic heavy ion collisions (HIC).
Modern experimental laboratories, such as at the Large Hadron Collider (LHC), the Relativistic Heavy Ion Collider (RHIC), the Super Proton Synchrotron (SPS), and the Heavy Ion Synchrotron (SIS), provide detailed measurements of hot strongly interacting matter that is excited in heavy ion collisions. 
On the other side, the phenomenology of QCD matter is accompanied by astrophysical observations of compact stars which provide insights into physics of cold but dense strongly interacting matter. The energy densities in both heavy ion collisions and the interiors of compact stars are of the same order and provide cold and hot scenarios of dense strongly interacting matter. A recently emerged branch of observational astronomy, the gravitational wave astronomy, has established a unique interplay between hot dynamical accelerator based phenomena of heavy ion collisions and the physics of cold static neutron stars. The gravitational waves emitted in general relativistic neutron star~(NS) collisions, like the so far only one detected GW170817~\cite{Abbott:2016blz}, allows to study the dynamics of NS mergers where due to dynamical compression temperatures up to 100 MeV can be reached~\cite{Most:2018eaw,Bauswein:2018bma}. The QCD input into NS mergers is the equation of state (EoS) which provides a relation between density, temperature and pressure. The particle content and their interactions provide a basis to construct an EoS so that a relation to the thermodynamical quantities can be obtained. Even though, the particle content is known, the relevance of different particles in different regimes of QCD is not known and should be controlled by effective interactions, details of which are not well established so far.

On the theoretical side there is no complete knowledge on the macroscopical properties of QCD matter. It is established that at low and moderate temperatures and densities the matter can be modeled as a gas of hadrons and resonances, while it is theorized that at higher temperatures and densities deconfinement occurs, a transition to quark and gluon degrees of freedom. Still, details of this transition are disputed, though some of them are well studied on both a theoretical~\cite{Shuryak:1980tp,McLerran:2007qj,Borsanyi:2013bia,Bazavov:2017tot} and experimental basis. It is still not known if deconfinement induces a thermodynamic phase transition, this question is still open, actively discussed, and considered as a main problem of QCD.

Along with the already extensively studied heavy ion collisions, NS mergers provide an unprecedented facility to study the dynamics of nuclear, hadronic, and possible quark matter that are excited in systems with spatial dimensions of order of kilometers. Though the matter content in both relativistic collision scenarios is very similar, there are differences due to strong gravitational effects in the interiors of NSs and in products of their mergers. 

This contribution is an attempt to show how the current knowledge of QCD phenomenology can be combined in a single framework for a QCD EoS applicable for laboratory-driven and astrophysical phenomena.

\pagebreak

\section{\label{sec:model}An approach to QCD equation of state:\\
Chiral SU(3)-flavor parity-doublet Polyakov-loop quark-hadron mean-field model}

\begin{figure}[h!]
  \centering
  \includegraphics[width=0.75\textwidth]{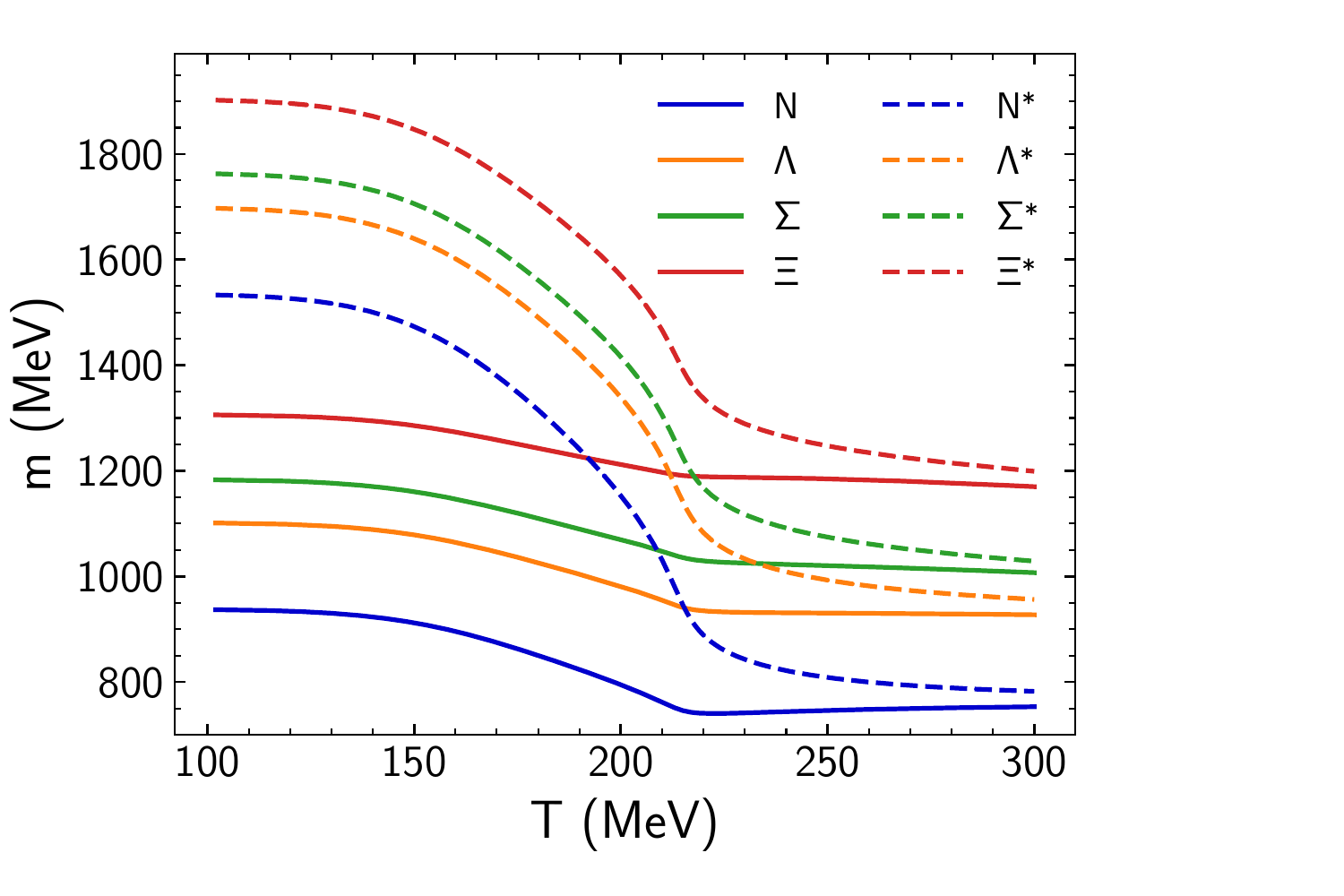}
  \caption{Effective masses of octet baryons (solid) and their respective parity partners (dashed) as functions of temperature $T$ obtained from the CMF model at $\mu_B=0$ for isospin symmetric matter, $Q/B=\frac12$. With the increase of the temperature the degeneracy in mass between baryons and respective parity partners arises.}
  \label{fig:masses}
\end{figure}

The Chiral SU(3)-flavor parity-doublet Polyakov-loop quark-hadron mean-field model~\cite{Steinheimer:2010ib,Motornenko:2019arp}, CMF, is a modification of the  $\sigma$-$\omega$ model to include quark degrees of freedom along with parity doubling for baryons~\cite{Detar:1988kn,Hatsuda:1988mv,Papazoglou:1996hf,Papazoglou:1997uw, Papazoglou:1998vr,Sasaki:2010bp,Steinheimer:2010ib, Steinheimer:2011ea,Dexheimer:2012eu,Mukherjee:2016nhb, Motornenko:2018hjw,Motornenko:2019arp}.  The CMF model is an attempt to include known phenomenological QCD interactions within a unified approach to describe interacting hadron-quark matter. An essential feature of the model is the chiral symmetry restoration for baryons, through the parity doubling for baryon octet states, so when chiral symmetry is restored they become degenerate in mass with respective parity partners~\cite{Aarts:2017rrl,Aarts:2018glk}. Within the CMF the same chiral fields generate dynamical masses for quarks, so the chiral condensate is a proxy interaction between hadrons and quarks.

\begin{figure}[h!]
  \centering
  \includegraphics[width=0.75\textwidth]{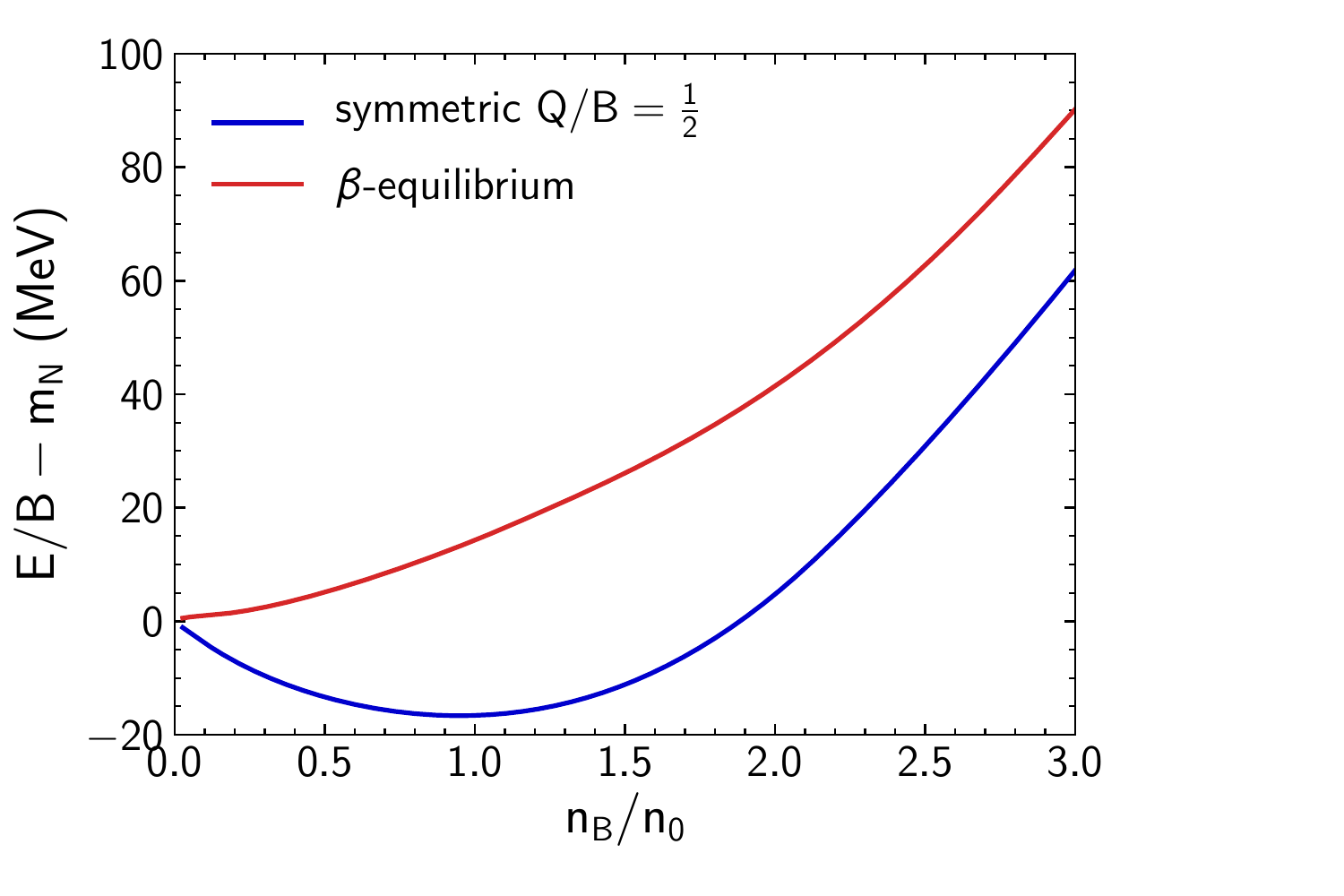}
  \caption{The CMF energy per baryon $E/B-m_N$ as function of baryon density $n_B$ normalized to nuclear saturation density $n_0=0.16$ fm$^{-3}$, blue -- for symmetric nuclear matter ,$Q/B=\frac12$, red -- for nuclear matter in $\beta$-equilibrium, total electric charge is zero $Q=0$ that is allowed by presence of leptons.}
  \label{fig:ground}
\end{figure}

The coupling parameters of the baryonic sector in the model are tuned to describe empirical data on cold nuclear matter~\cite{Papazoglou:1998vr}. At $T=0$ the model produces nuclear ground state with the following properties: 
ground state density $n_0 = 0.16~({\rm fm}^{-3})$, binding energy per nucleon is $E/B = -15.2$ (MeV), asymmetry
energy $S_0 = 31.9$ (MeV), and compressibility $K_0 = 267$ (MeV). The energies per baryon for nuclear matter at $T=0$ are presented in Fig.~\ref{fig:ground}, where the ground state of isospin symmetric matter is presented along with the results for neutron rich matter in $\beta$-equilibrium. The CMF model can be used to provide the EoS for electric neutral  neutron-rich matter in $\beta$-equilibrium, where leptons, electrons and muons, are sufficient to compensate the electric charge of hadrons, this is the matter present in interiors of NSs.

\begin{figure}[h!]
\centering
\includegraphics[width=.49\textwidth]{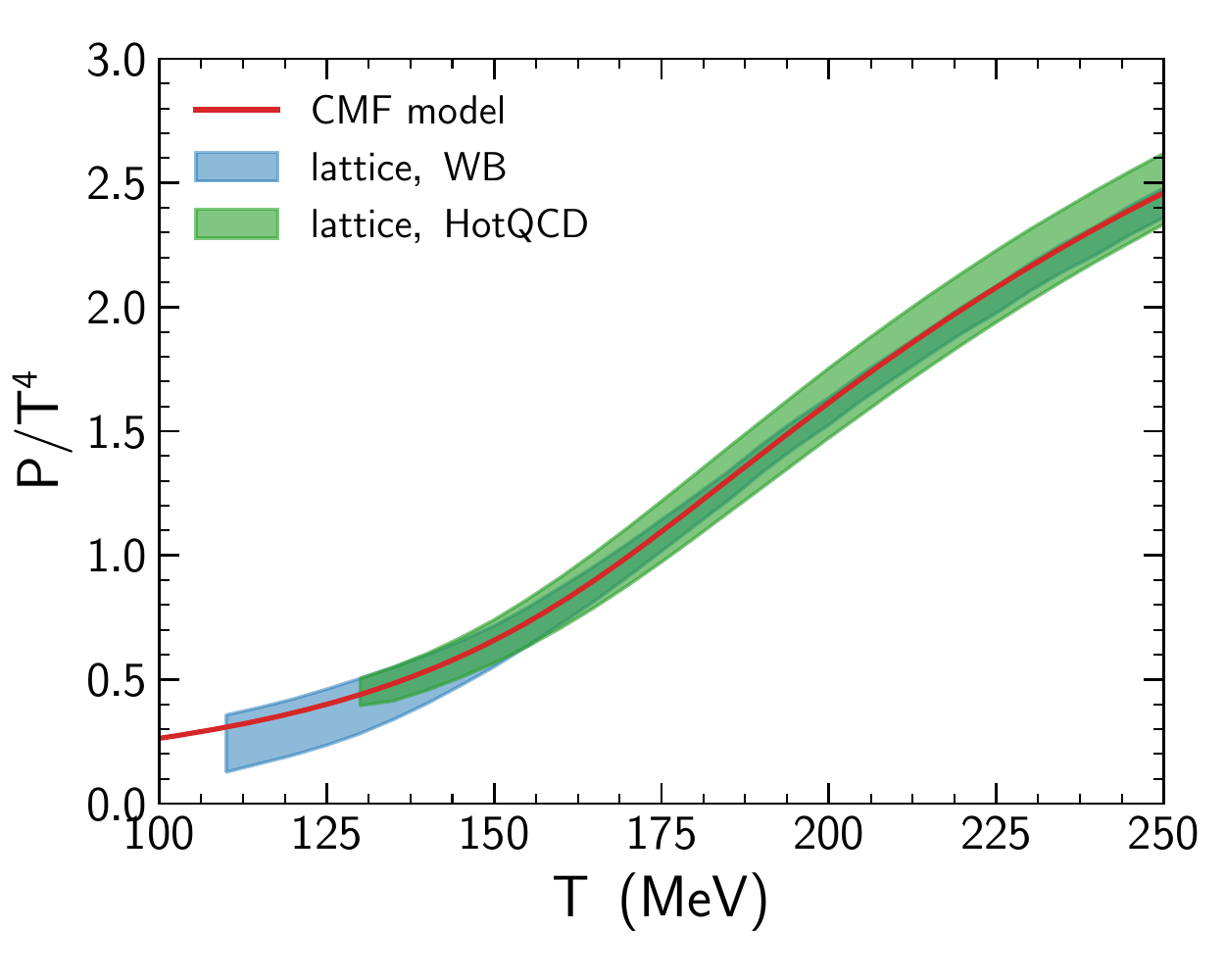}
\includegraphics[width=.49\textwidth]{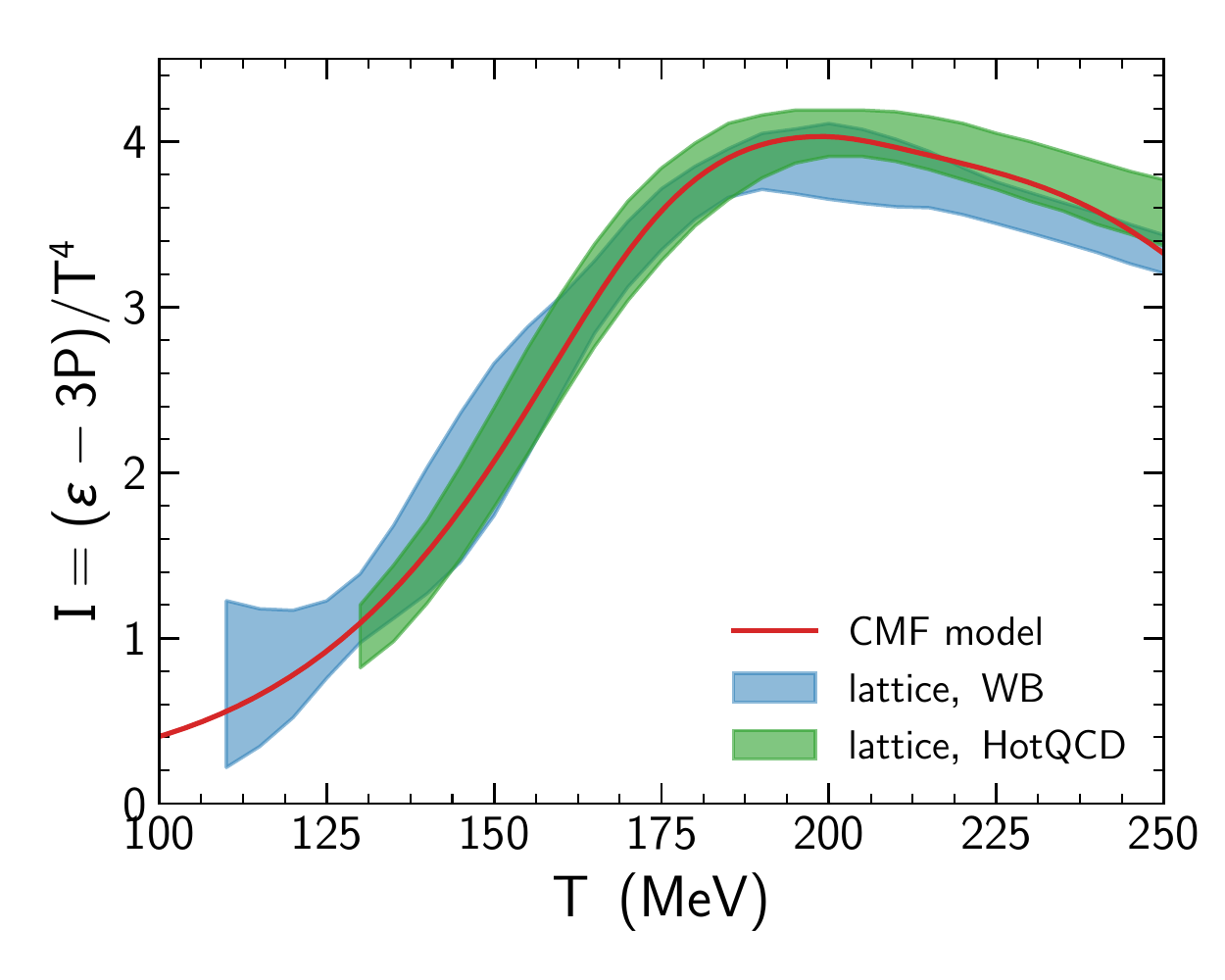}
\caption{ Pressure $P$ $I$ ({\bf right}) and trace anomaly $I$ ({\bf left}) at $\mu_B=0$ as function of temperature $T$. Comparison between the CMF model predictions and LQCD results \cite{Borsanyi:2013bia, Bazavov:2014pvz}.}
\label{fig:fit}
\end{figure}

\begin{figure}[h!]
  \centering
  \includegraphics[width=0.49\textwidth]{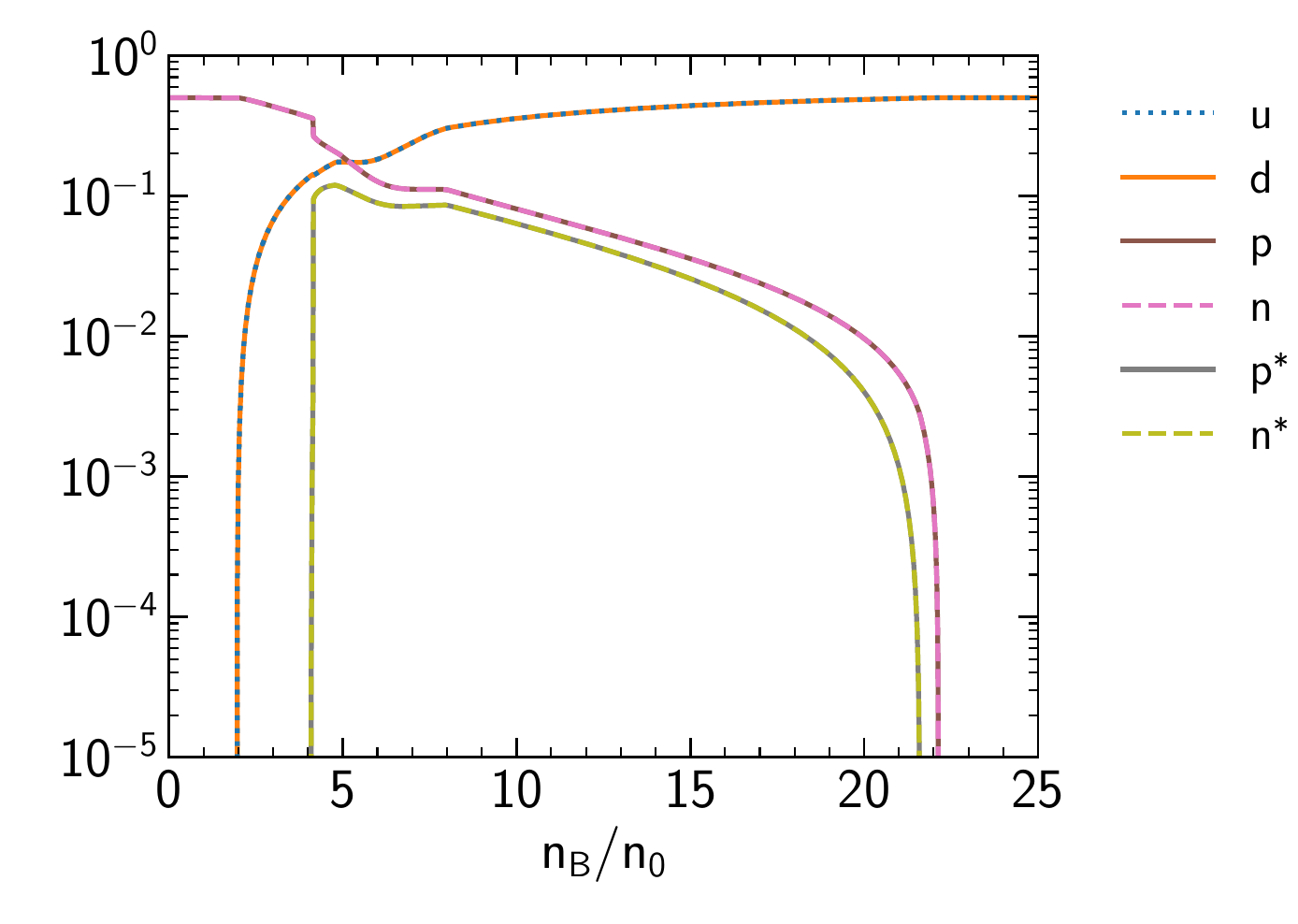}
  \includegraphics[width=0.49\textwidth]{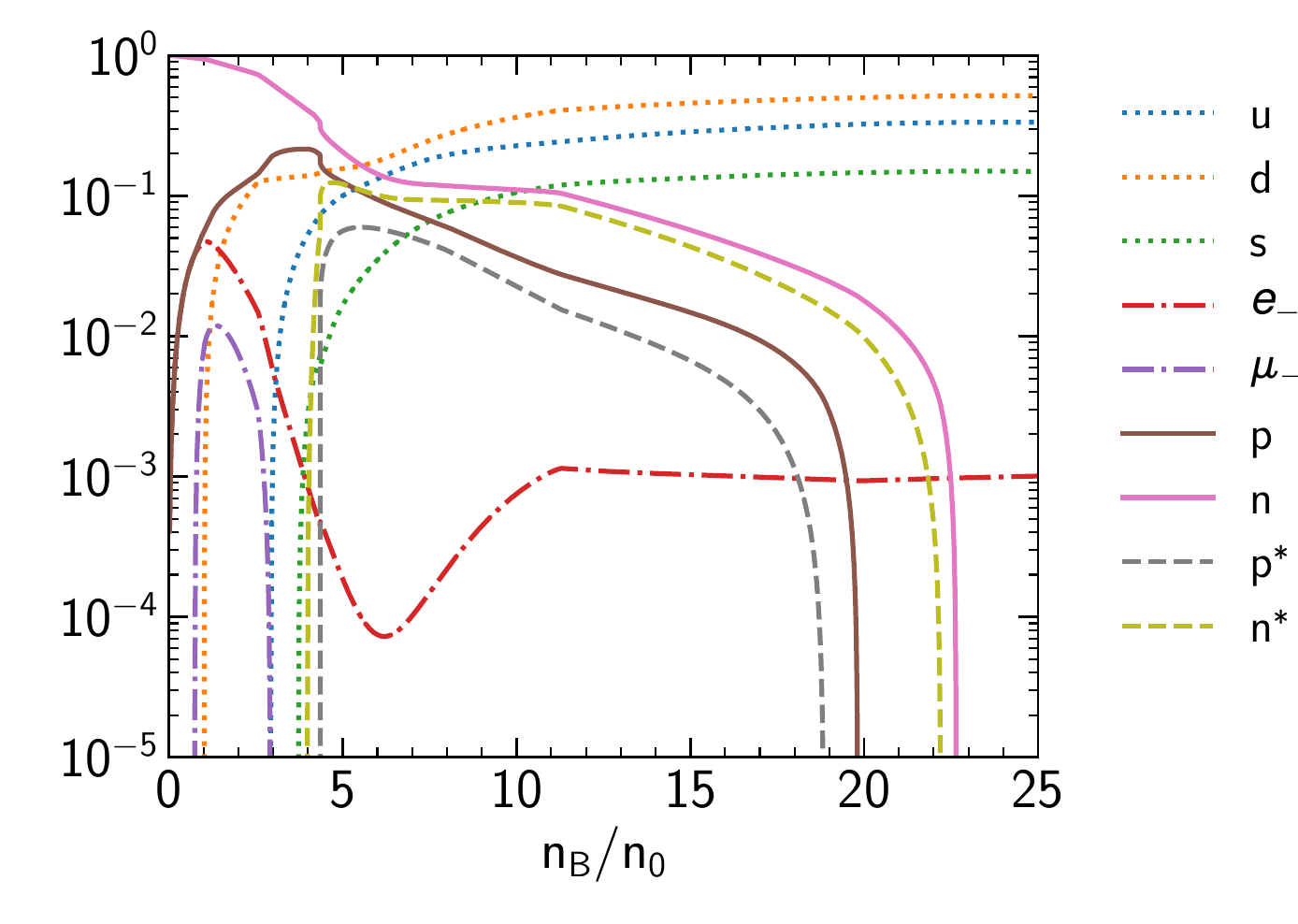}
  \caption{Particle density ratios to the density of baryons $n_i/n_B$ at $T=0$, for quarks a factor of $1/3$ is used, presented as functions of baryon density $n_B$. The CMF-results are obtained for isospin symmetric matter ({\bf left}) and in $\beta$-equilibrium ({\bf right}).}
  \label{fig:content}
\end{figure}

The hadronic sector of the CMF model is based on a $\sigma$-$\omega$ model in mean-field approximation. Mesonic mean fields provide interactions between baryons, namely baryons from the SU(3)$_f$ baryon octet along with their parity partners, i.e. states with the same quantum numbers but opposite parity. To provide the dynamical mass generation, the octet baryons and their partners are coupled to scalar chiral fields $\sigma$ and $\zeta$, non-strange and strange chiral condensates, respectively. The $\sigma$ and $\zeta$ fields serve as order parameters for the chiral transition:
\begin{eqnarray}
m^*_{\text{i}\pm} = \sqrt{ \left[ (g^{(1)}_{\sigma \text{i}} \sigma + g^{(1)}_{\zeta \text{i}}  \zeta )^2 + (m_0+n_\text{s} m_\text{s})^2 \right]} \pm  \left(g^{(2)}_{\sigma \text{i}} \sigma + g^{(2)}_{\zeta \text{i}} \zeta\right) ~.
\label{effmass}
\end{eqnarray}
Here $+$ is used for positive and $-$ is for negative parity states, $g^\text{(j)}_\text{i}$ are the coupling constants of baryons to the two scalar fields, $m_0=759$ MeV is the baryon mass at the restored phase. In addition, there is an SU(3)$_f$ symmetry-breaking mass term proportional to the strangeness content of the baryons, where $n_s$ is the number of strange quarks in the baryon, and $m_s=130$ MeV is the mass of the strange quark. The couplings $g^\text{(j)}_\text{i}$ are tuned to reproduce the vacuum masses of baryons. Figure~\ref{fig:masses} presents how the mass degeneracy is restored with increasing temperature $T$ at vanishing chemical potential, $\mu_B=0$. At lower temperatures there is a significant mass gap between partners which diminishes as chiral symmetry restores.

The values of the mesonic fields are driven by the thermal potentials of quarks and by the scalar meson interaction that provides spontaneous chiral symmetry breaking:
\begin{equation}
V = V_0 + \frac{1}{2} k_0 I_2 - k_1 I_2^2 - k_2 I_4 + k_6 I_6 ~,
\label{veff}
\end{equation}
with
\begin{eqnarray}
    I_2 = (\sigma^2+\zeta^2),~~ I_4 = -(\sigma^4/2+\zeta^4), ~~
    I_6 = (\sigma^6 + 4\, \zeta^6)
\end{eqnarray}
where $V_0$ is fixed by requiring a vanishing potential in vacuum.\\ 

The quark sector of the model is implemented similarly to the PNJL approach~\cite{Fukushima:2003fw}. 
The thermal contribution of quarks is driven by the value of Polyakov loop $\Phi$, which plays the role of the order parameter for the deconfinement transition.
Quark thermal contribution is coupled to the Polyakov loop $\Phi$ and their grand canonical potential $\Omega$ is expressed as
\begin{equation}
	\Omega_{q}=-T \sum_{i\in Q}{\frac{d_i}{(2 \pi)^3}\int{d^3k \ln\left(1+\Phi \exp{\frac{-\left(E_i^*-\mu^*_i\right)}{T}}\right)}}
\end{equation}
and
\begin{equation}
	\Omega_{\overline{q}}=-T \sum_{i\in Q}{\frac{d_i}{(2 \pi)^3}\int{d^3k \ln\left(1+\Phi^* \exp{\frac{-\left(E_i^*+\mu^*_i\right)}{T}}\right)}}.
\end{equation}
The sums run over all light quark flavors (u, d, and s), $d_i$ is the corresponding degeneracy factor, $E^*_i=\sqrt{m^{*2}_i+p^2}$ is the energy. The quark chemical potential $\mu^*$ is defined by the quarks quantum numbers and is not modified by repulsive interactions as those are disfavored by lattice QCD (LQCD) calculations~\cite{Steinheimer:2010sp,Steinheimer:2014kka}.  Note that two- and three- quark contributions to $\Omega$ are omitted in the CMF model since hadronic excitations are explicitly included.

The effective masses of the light quarks are generated by the $\sigma$ and $\zeta$, non-strange and strange chiral condensates respectively. The quark explicit mass terms $\delta m_q=5$ MeV, and $\delta m_s=150$ MeV for the strange quark, and $m_{0q}=253$ MeV correspond to an explicit mass term which does not originate from chiral symmetry breaking:
\begin{eqnarray}
m_{q}^* & =-g_{q\sigma}\sigma+\delta m_q + m_{0q},&\nonumber\\
m_{s}^* & =-g_{s\zeta}\zeta+\delta m_s + m_{0q}.&
\end{eqnarray}

The dynamics of the Polyakov-loop is controlled by the effective Polyakov-loop potential $U(\Phi, \Phi^*, T)$~\cite{Ratti:2005jh}:
\begin{align}
	U=-\frac12 a(T)&\Phi\Phi^*
	+ b(T)\log[1-6\Phi\Phi^*+4(\Phi^3 + \Phi^{*3})-3(\Phi\Phi^*)^2],\nonumber\\
    a(T)&=a_0 T^4+a_1 T_0 T^3+a_2 T_0^2 T^2,~
    b(T)=b_3 T_0^4
\end{align}

The parameters of this potential are fixed to the lattice QCD data on the interaction measure $I=\frac{\varepsilon-3 P }{T^4}$~\cite{Motornenko:2019arp}. Figure~\ref{fig:fit} shows a comparison between the CMF predictions for $\mu_B=0$ thermodynamics and LQCD data~\cite{Borsanyi:2013bia, Bazavov:2014pvz}. The parameter values of $U(\Phi, \Phi^*, T)$ ensure the applicability of the CMF model at high temperatures and make a fair description of LQCD data possible within one unified approach for both hot QCD thermodynamics and cold nuclear matter description.

Figure~\ref{fig:content} demonstrates the particle content at $T=0$ for isospin symmetric matter and for matter in $\beta$-equilibrium.  At $n_B\approx4\,n_0$ the phase transition associated with chiral symmetry restoration takes place. At that region parity partners become degenerate in mass so the degeneracy of nucleons is effectively increased by a factor of two. Note that at $T=0$ baryons are presented only by nucleons and their parity partners. Though heavier baryons and their resonances are present in the model, but at $T=0$ they are suppressed by the strong repulsion and they start to appear only with increasing temperature. After chiral symmetry restoration a significant fraction of quarks is present, so within the CMF model there's no phase transition that separates the quark and hadronic phases. Only at very high densities $n_B\gtrsim22\,n_0$ the content of the matter is composed by only quarks.

To mimic hard-core repulsive interactions among baryons, an excluded-volume effect is incorporated within the CMF model~\cite{Steinheimer:2011ea}. This approach ensures that  quarks and gluons dominate at high densities and temperatures. This results in all particle densities, including quarks, being reduced, as parts of the system are occupied by EV-hadrons:
\begin{eqnarray}
\rho_j=\frac{\rho^{\rm id}_j (T, \mu^*_j - v_j\,p)}{1+\sum\limits_i ^{\rm HRG} v_i \rho^{\rm id}_i (T, \mu^*_i - v_i\,p)},
\end{eqnarray}
where the $v_i$ are the eigenvolume coefficients for different species. $p$ is pressure of the system without the contribution of mean fields, $\mu_i^*$ denotes the modified chemical potential of the hadrons.
The $v$ are fixed to $v_B = 1$~fm$^3$ for (anti-) baryons, $v_M = 1/8$~fm$^3$ for mesons, and are set to zero $v_q = 0$ for quarks, where indices $i$ and $j$ run through all states present in the hadron-resonance gas and, additionally, quarks.

\section{\label{sec:ph0diag}The CMF model's QCD phase diagram}

\begin{figure}[h!]
\centering
\includegraphics[width=.75\textwidth]{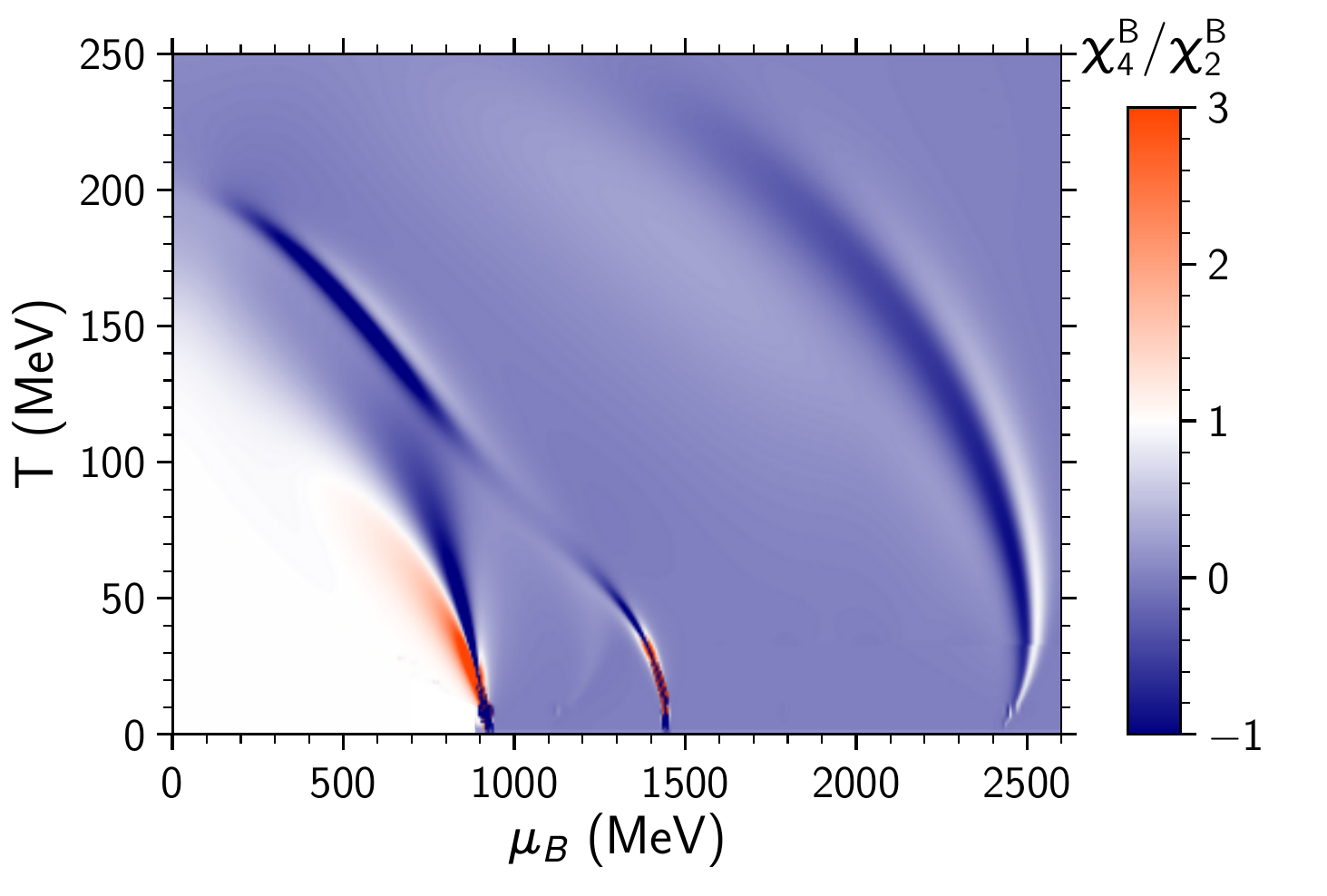}

\caption{CMF baryon number kurtosis, ratio of fourth and second order susceptibilities, $\chi_4^{\rm B}/\chi_2^{\rm B}$  in the baryon chemical potential and temperature, $\mu_{\rm B}-T$, plane. The three distinctive critical regions and their remnants spread from $T=0$ up to $T>200$ MeV.}
\label{fig:suscept-diag_muT}
\end{figure}

Numerous phenomenological features of the CMF model suggest a rather rich phase structure of the model. The critical phenomena associated with the nuclear liquid-vapor phase transition, chiral symmetry restoration and quark appearance, present in the CMF model provide a complicated interplay. To determine the location of different phases the baryon number susceptibilities $\chi_n^B$ are used:
\begin{eqnarray}
\chi_n^B= \frac{\partial^n (P/T^4)}{(\partial \mu_B/T)^n}\,,
\label{eq:susc}
\end{eqnarray}
Higher-order baryon number susceptibilities are measures of baryon number fluctuations in the grand canonical ensemble. $\chi_n^B$ rise proportionally to the increasing powers of the correlation length \cite{Stephanov:2008qz}, so an increase in the correlation length is reflected in large values of the 2nd and higher-order susceptibilities in the vicinity of the critical point and in the region of a phase transition or crossover. Hence, these quantities are useful indicators of critical behavior. Deviations of $\chi_n^B$ from the corresponding baselines indicate a transformation between different phases, which is reflected usually in a non-monotonic behavior of these observables, e.g. kurtosis $\chi_4^B/\chi_2^B$.

The CMF model kurtosis, $\chi_4^B/\chi_2^B$,~Fig.~\ref{fig:suscept-diag_muT}, indicates several structures in the $\mu_B-T$ plane. Those structures are attributed to the thermal behavior of various order parameters that experience a rapid change in these regions. 

\begin{figure}[h!]
  \centering
  \includegraphics[width=0.75\textwidth]{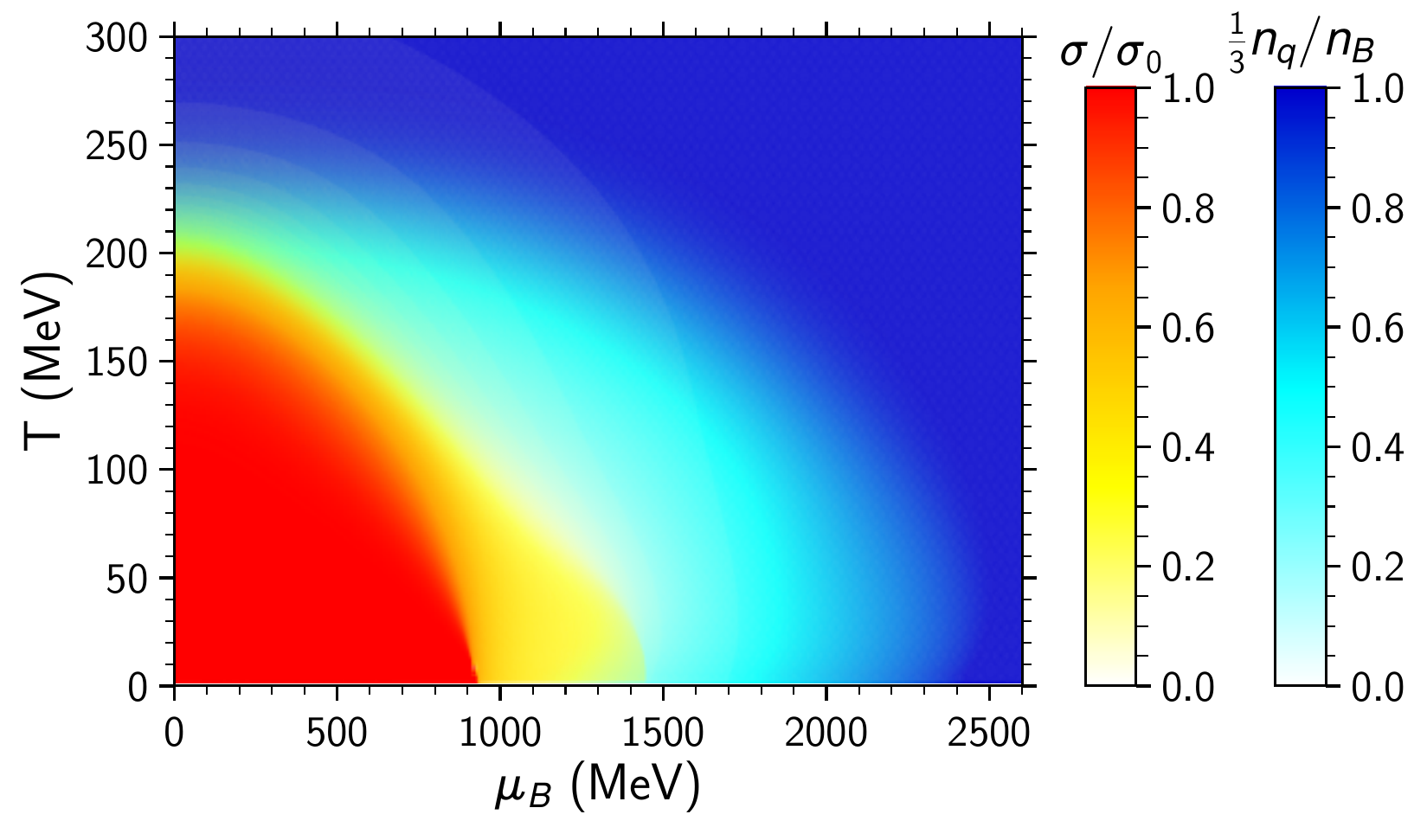}
  \caption{Colormaps for chiral condensate $\sigma$ and quark fraction $\frac13 n_q/n_B$ along the $\mu_B-T$ plane for isospin symmetric matter. Note that significant quark fraction appears only after chiral symmetry is restored, so pure quark matter arises at high $\mu_B$ or $T$. Mind the transparency of white color, so two quantities are presented on the same plot.}
  \label{fig:q-sigma}
\end{figure}

At low and intermediate temperatures and chemical potentials rapid growth in baryonic density provides a change in value of kurtosis' baseline from 1 to 0. This transition is associated with the nuclear liquid-vapor phase transition,  which, at low temperatures, is a phase transition from a dilute gas of nucleons to dense nuclear matter. This transition smoothens out to a crossover at $T\gtrsim20$ MeV, so the change in baryon density is smooth and not step-like. Note, even at $\mu_B=0$ and $T>100$ MeV the remnants of nuclear liquid-vapor transition are significant and still contribute to the critical behavior of the CMF model.

At higher temperatures and chemical potentials the second consecutive critical region is attributed to chiral symmetry restoration. This transition is driven by the rapidly vanishing chiral condensate $\sigma$, see Fig.~\ref{fig:q-sigma}, reflecting a degeneracy among parity partners and decreasing quark masses. This transition is of a first order with a critical temperature $T_{\rm CP}^{\rm chiral}\approx20$ MeV , so the chiral condensate drops discontinuously at $T<T_{\rm CP}^{\rm chiral}$, at higher temperatures this transition is of crossover-type. Only after the chiral symmetry restoration the quark fraction is significant in the CMF model. The free quark contribution to baryon density, $\frac13n_q/n_B$, is presented in Fig.~\ref{fig:q-sigma}.

Quarks in the model appear smoothly, so there is no deconfinement related phase transition. Quarks appear in the chirally restored phase and they are in mixture with other baryons and mesons. This smooth appearance of free quarks and their presence along with other baryons is similar to a quarkyonic matter picture~\cite{McLerran:2007qj, McLerran:2018hbz}, where quarks appear from Fermi sea and Baryons reside on the Fermi shell.

Only at sufficiently high temperatures or large baryon densities the content of the matter is comprised only by quarks and gluons within the Polyakov loop potential. This transition produces a third consecutive region in kurtosis in the $\mu_B-T$ plane. This smooth transition is driven by the reduction of degrees of freedom, as hadrons are no longer contributing to the thermodynamics. 

\section{\label{sec:par}Probing the phase diagram by heavy ion collisions}

During its expansion the system created in a heavy-ion collision probes various regions of the phase diagram. The trajectory of the system evolution can be estimated by hydrodynamic modeling. The input to hydro simulations is the initial stage produced in a rapid violent nuclear collision. In the initial state the entropy is produced by a violent shock compression~\cite{Stoecker:1986ci}. 
While the system cools down during expansion, the entropy increases only moderately due to a rather small viscosity~\cite{Csernai:2006zz,Romatschke:2007mq}, thus an isentropic expansion scenario is a reasonable approximation \cite{Steinheimer:2007iy}. 

We illustrate which regions of the phase diagram are probed in heavy ion collisions at low and moderate energies by using the one dimensional Taub adiabat model~\cite{Taub:1948zz,Stoecker:1978jr,Thorne:taub_adiabat}. This model describes the expansion by lines of constant entropy per baryon $S/B=\,$const (isentropes).
These lines describe the isentropic matter evolution of ideal fluid dynamics at different collision energies.

The isentropic expansion of the equilibrated matter continues until the system becomes so dilute that the chemical as well as the kinetic freeze-out occur when the chemical composition is fixed.

The calculation of initial entropy per baryon ($S/A$) is done assuming a 1-dimensional stationary scenario of central HIC -- two colliding slabs of cold nuclear matter \cite{Baumgardt:1975qv,Stoecker:1978jr,Stoecker:1980uk, Stoecker:1981za, Stoecker:1981iw,Hahn:1986mb, Stoecker:1986ci}. The relativistic Rankine-Hugoniot equation (Taub adiabat), RRHT,~\cite{Taub:1948zz, Thorne:taub_adiabat} provides conservation of the baryon number, energy and momentum across the shock front. Using the RRHT one can directly associate entropy  with the collision energy. The RRHT-equation provides thermodynamic properties across the shock front 
\begin{equation}
    \label{eq:Taub}
    (P_0+\varepsilon_0)\, (P+\varepsilon_0)\, n^2=(P_0+\varepsilon)\, (P+\varepsilon)\, n^2_0\,,
\end{equation}
where $P_0$, $\varepsilon_0$ and $n_0$ correspond to the initial pressure, energy density, and baryon density in the local rest frame of each of the two slabs. The two symmetric slabs consist of the nuclear matter in the ground state, $P_0=0,~\varepsilon_0/n_0 - m_N=-16$ MeV and $n_0=0.16~ {\rm fm^{-3}}$. The created density is related to the collision energy as:
\begin{equation}
    \label{eq:stopping}
    \gamma^{\rm CM}=\frac{\varepsilon n_0}{\varepsilon_0 n},~\gamma^{\rm CM}=\sqrt{\frac{1}{2}\left(1+\frac{E_{\rm lab}}{m_N}\right)}\,.
\end{equation}
Here $\gamma^{\rm CM}$ is the Lorentz gamma factor in the center of mass frame of the HIC and $E_{\rm lab}$ is the beam energy per nucleon in the laboratory frame of a fixed target collision.
This relation can be obtained from the full stopping condition~\cite{Stoecker:1978jr,Stoecker:1980uk, Stoecker:1981za, Stoecker:1981iw,Hahn:1986mb, Stoecker:1986ci, 1103.3988}. 
The initial state thermodynamics (density, temperature and entropy) of the hot, dense participant matter is obtained from~\cref{eq:Taub,eq:stopping} as a function of the collision energy. The known initial entropy yields the lines of constant entropy which leads to the trajectories of the heavy ion collisions in the phase diagram.

\begin{figure}[h!]
\centering
\includegraphics[width=.49\textwidth]{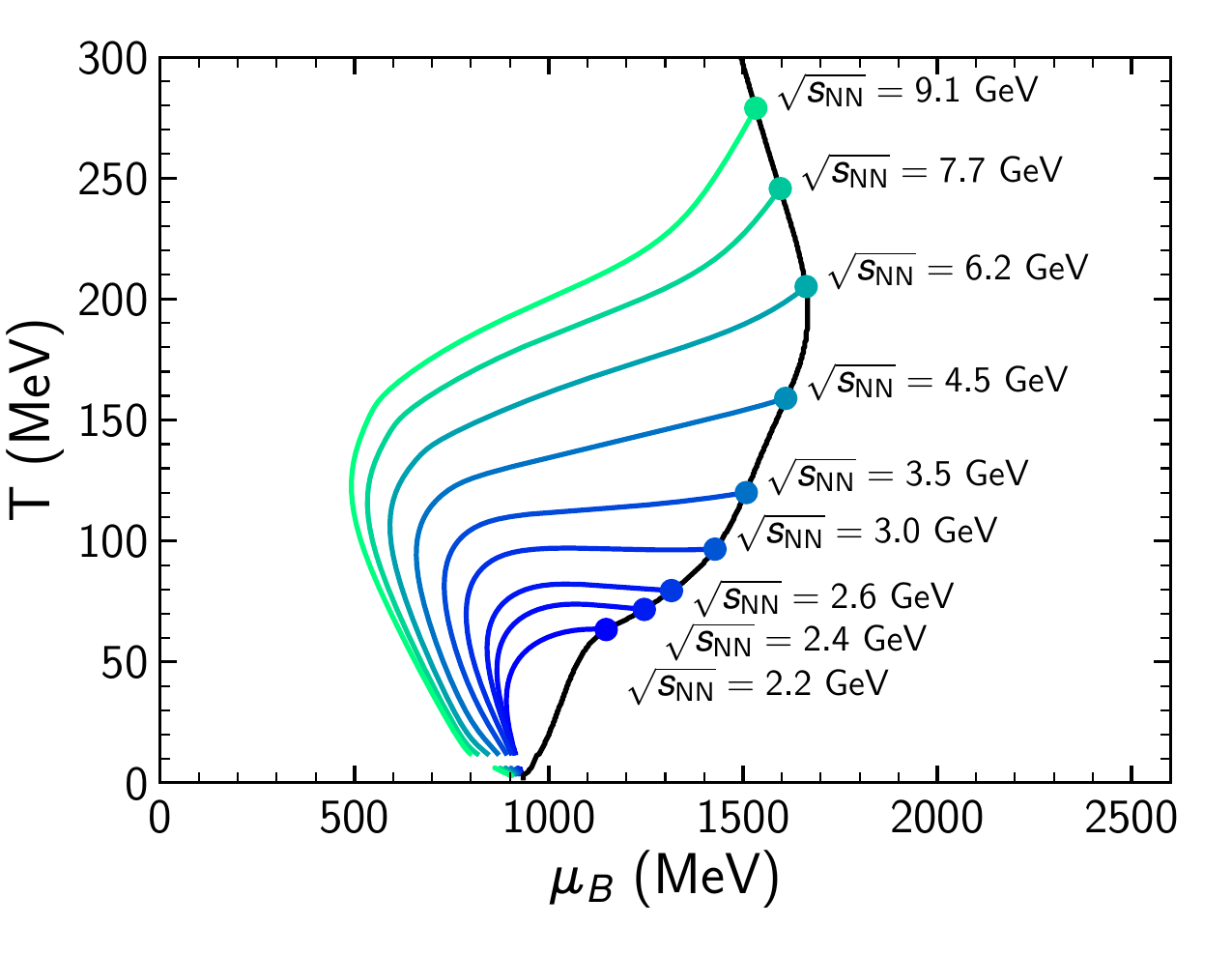}
\includegraphics[width=.49\textwidth]{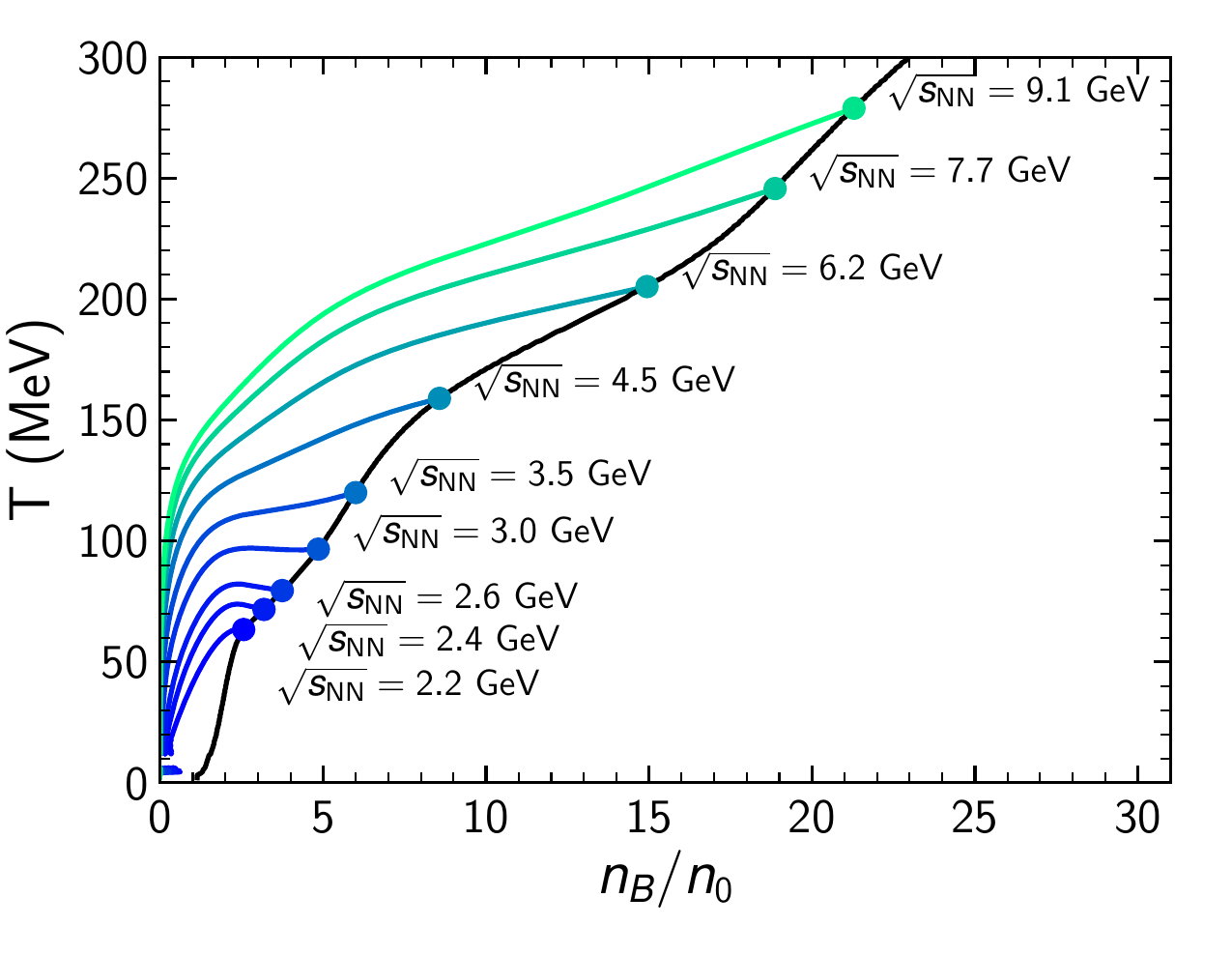}

\caption{Evolution for different collision energy $\sqrt{s_{\rm NN}}$ of system excited in heavy-ion collisions along the $\mu_B-T$ ({\bf left}) and $T-n_B$ ({\bf right}) phase diagram. Initial state as implicit function of $\sqrt{s_{\rm NN}}$ is calculated by Taub adiabat and is presented by the black line. Colored lines -- isentropic lines of constant entropy per baryon $S/A$ at different bombarding energies $\sqrt{s_{NN}}$ respectively.}
\label{fig:isentropes}
\end{figure}

The predicted isentropic expansion trajectories are shown in the $\mu_B-T$ phase diagram in~\cref{fig:isentropes}. 

The RRHT-adiabat scenario predicts a very strong compression and heating already at intermediate lab (fixed target) bombarding energies. The hot and dense system passes the chiral transition predicted by the present CMF model already at $E_{\rm lab}\approx 2$~A\,GeV, i.e. at energies available at GSI's SIS18 accelerator facility. Here the specific total entropy is predicted to reach $S/A\approx 3$, in accord with previous RMF-calculations~\cite{Hahn:1986mb} which also used the 1-D RRHT-scenario. The $\mu_B-T$ values, $T\approx 70$ MeV, $\mu_B\approx 1.2$ GeV, with net baryon densities $n_B/n_0\approx 3$, reached here in HIC, coincide with the $\mu_B-T$ values reached in binary NS collisions,  as recent general relativistic fully 3+1-dimensional megneto-hydrodynamical calculations have confirmed~\cite{Hanauske:2017oxo,Most:2018eaw} for the gravitational wave event GW170817. At these temperatures and densities, $T\approx 70$ MeV and $n_B/n_0\approx 3$, the RRHT model predicts that there are about 20\% of the dense matter are already transformed to quarks.

Heavy ion fixed target experiments of SIS at FAiR and SPS at CERN as well as STAR BES program at RHIC probe temperatures from $50<T<280$ MeV and chemical potentials from $500<\mu_B<1700$ MeV for the collision energy range $\sqrt{s_{\rm NN}}<10$ GeV considered here. 
In this region the CMF model shows not an additional phase transition, but the remnants of the nuclear liquid-vapor transition at $T \approx 20$ MeV. The chiral transition at larger chemical potentials can influence the dynamical evolution, too. 
The present results suggest that heavy-ion collisions mostly probe regions where the nuclear matter liquid-vapor critical point dominates -- hence, the observed baryon fluctuations are largely due to remnants of the nuclear liquid-vapor phase transition.  
This had been suggested also in previous works  \cite{Fukushima:2014lfa,Mukherjee:2016nhb,Vovchenko:2016rkn,Vovchenko:2017ayq,Steinheimer:2018rnd,Ye:2018vbc}.
The CP associated with the chiral symmetry restoration in the CMF model lies at $\mu_B \approx 1.5$ GeV and $T \approx 20$ MeV. 
This high density region is, to the best of our knowledge, reachable in the interior of NS and in binary general relativistic NS mergers \cite{Dietrich:2015iva,Radice:2016rys,Most:2018eaw,Bauswein:2018bma,Hanauske:2019qgs,Hanauske:2019new}.

\section{\label{sec:susc}QCD equation of state and neutron stars}
 Interiors of NSs consist of matter that surpasses nuclear saturation density by substantially large factors. The theory for these large densities is not complete and details of the equation of state are missing. As discussed above, the situation is similar to the matter created in HIC where details of equation of state at high temperatures are not known. There are ongoing discussions of the relevance of hyperonic, quark, and hadronic degrees of freedom in cold and dense NS matter.

 The CMF model is a general purpose equation of state that  with the same set of parameters can be applied to describe the NS matter. To produce the equation of state for compact stellar objects the condition of electric neutrality is imposed so $\beta$-equilibrium is ensured, these constraints demand the presence of leptons. This changes the particle content as depicted at Fig.~\ref{fig:content}. As a result, strangeness and hypercharges obtain finite non-zero values and quark-hadron degrees of freedom are present as well. The temperatures in the star interiors are negligibly small compared to hot QCD scales, so the $T=0$ EoS can be applied.

The resulting CMF EoS at $T=0$ is used as input for the Tolman-Oppenheimer-Volkoff (TOV) equation, so a relation between the mass and the radius can be obtained for any static, spherical, gravitationally bound object \cite{Tolman:1939jz, Oppenheimer:1939ne}, i.e. here a static NS. The NS surface layers presumably consist of mostly neutron rich nuclei and clusters in chemical and $\beta$- equilibrium. Those nuclei are not yet a part of the CMF model. Therefore, we use a classical crust-EoS~\cite{Baym:1971pw} matched additionally to the CMF-EoS at $n_B\approx0.05~{\rm fm^{-3}}$. 
\begin{figure}[t]
\centering
\includegraphics[width=.49\textwidth]{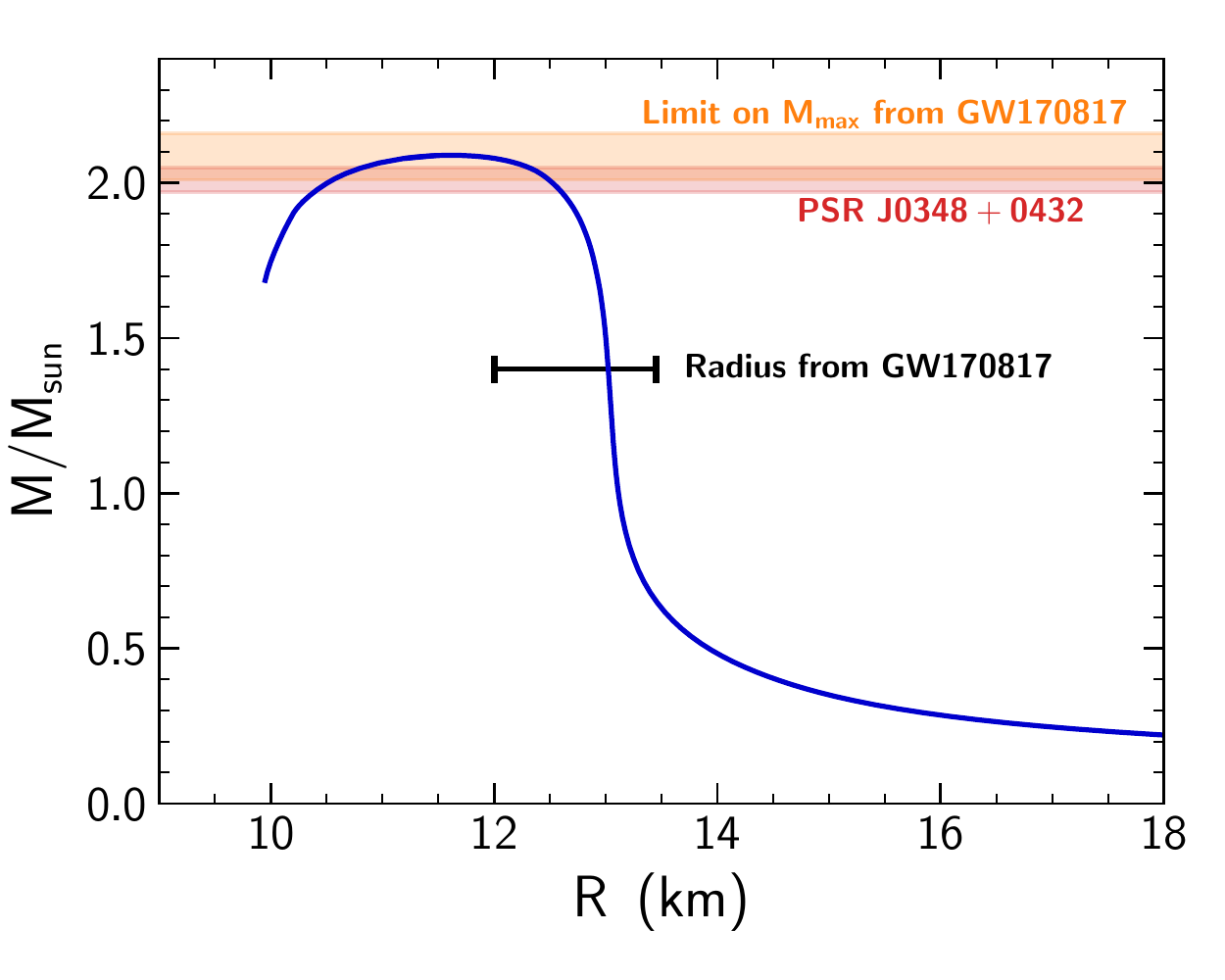}
\includegraphics[width=.49\textwidth]{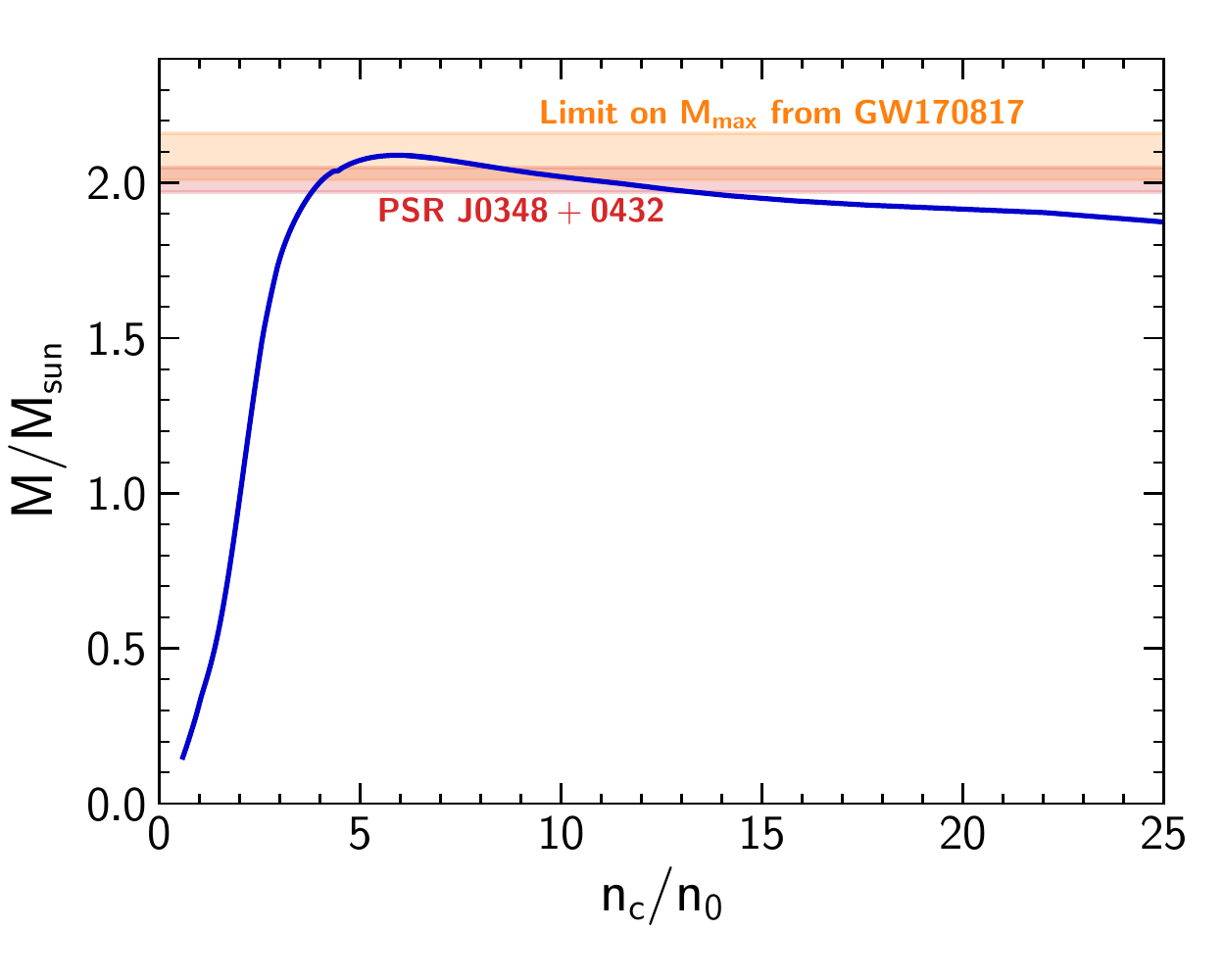}
\caption{Mass-radius relation ({\bf left}) and mass - central density relation ({\bf right}) for NSs calculated using the CMF equation of state. Yellow colorbands present constraint on the maximum mass of NS obtained in~\cite{Rezzolla:2017aly}, red colorband  -- the mass of NS PSR J0348+0432~\cite{2013ApJ...763...81L}.}
\label{fig:mr}
\end{figure}

The NS mass-radius relation obtained using the TOV equation with the CMF EoS is presented in Fig.~\ref{fig:mr}. The discussion of the quark content of the stars is presented in~\cite{Motornenko:2019arp}. Note that an unstable branch in the mass radius diagram is created by stars where the quark contribution to the baryon density is $30\%$ and more. The central densities of the stable stars can not exceed $n_B=6\,n_0$, as shown in the right panel of~\cref{fig:mr} where the maximum mass indicates the ``last stable star''. The continuous monotonous transition from NS matter to a deconfined quark phase provides a smooth appearance of quarks in the star structure and prevents a ``second family'' of stable solutions. Therefore there is no strict separation between a quark core and the hadronic interior of the star. This is a CMF result due to the Polyakov loop implementation of the deconfinement mechanism and no vector repulsion among quarks. Though LQCD data disfavors repulsive forces for quarks, there are currently active discussions concerning vector repulsion in dense baryonic matter in NS interiors~\cite{Benic:2014jia,Song:2019qoh}.

\begin{figure}[t]
\centering
\includegraphics[width=.49\textwidth]{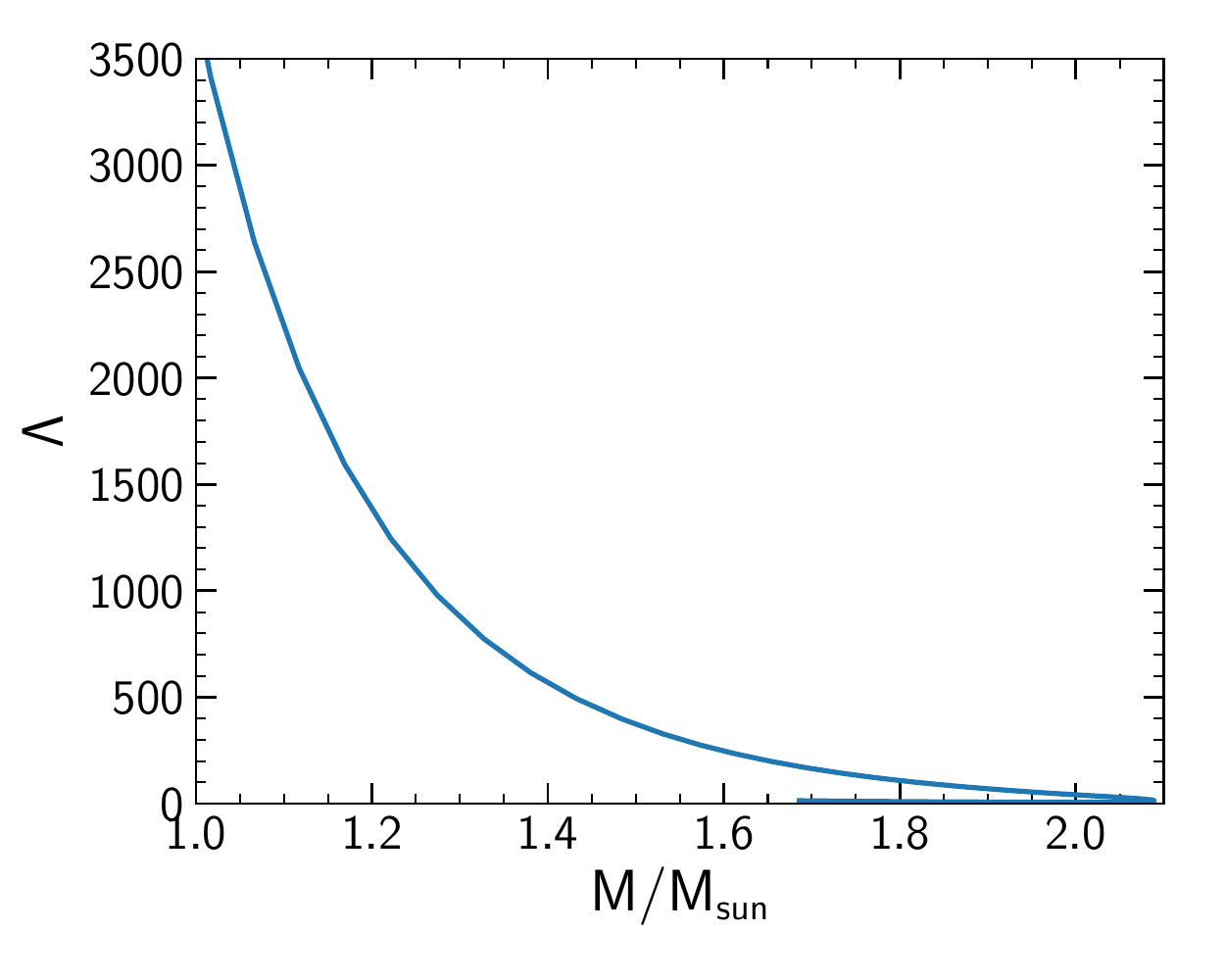}
\includegraphics[width=.49\textwidth]{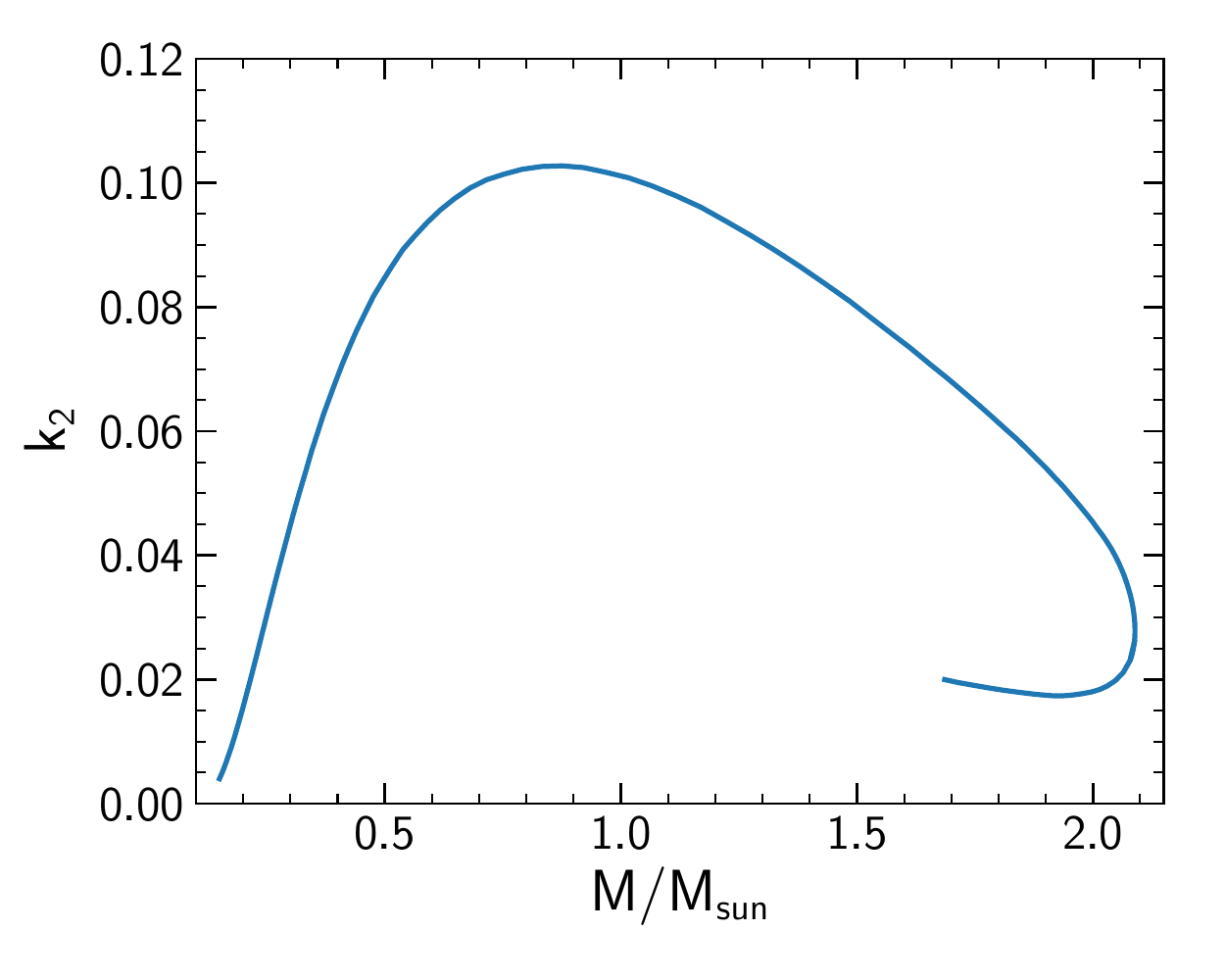}
\caption{The CMF tidal deformability $\Lambda$ ({\bf left}) and second Love number $k_2$ ({\bf right})  as functions of NS mass.}
\label{fig:lambda}
\end{figure}

During binary NS mergers the colliding participants experience significant non-spherical gravitational fields induced by their merge companions. These modifications of the gravitational field induce tidal deformations of the stars. The response of a NS to the non-spherical field depends strongly on the EoS and is reflected in the tidal deformability coefficient $\lambda$~\cite{0911.3535}. The tidal deformability $\lambda$ is a measure of the induced quadruple moment $Q_{ij}$ as a response to the external tidal field $\mathcal{E}_{ij}$:
\begin{equation}
    Q_{\ij}=-\lambda \mathcal{E}_{ij}\,.
\end{equation}
$\lambda$ is directly proportional to the second Love number $k_2$:
\begin{equation}
\lambda = \frac{2}{3}k_2 R^5\,.
\end{equation}
For convenience, usually the dimensionless tidal deformability $\Lambda$ is presented as:
\begin{equation}
    \Lambda = \frac{\lambda}{M^5}=\frac{2}{3}k_2\left(\frac{R}{M}\right)^5\,.
\end{equation}
Here, $M$ and $R$ are the mass and radius of the NS. We present the resulting dimensionless tidal deformability coefficient $\Lambda$ and second Love number $k_2$ in Fig.~\ref{fig:lambda}, see Ref.~\cite{Motornenko:2019arp} for discussion.

\begin{figure}[h!]
\centering
\includegraphics[width=.65\textwidth]{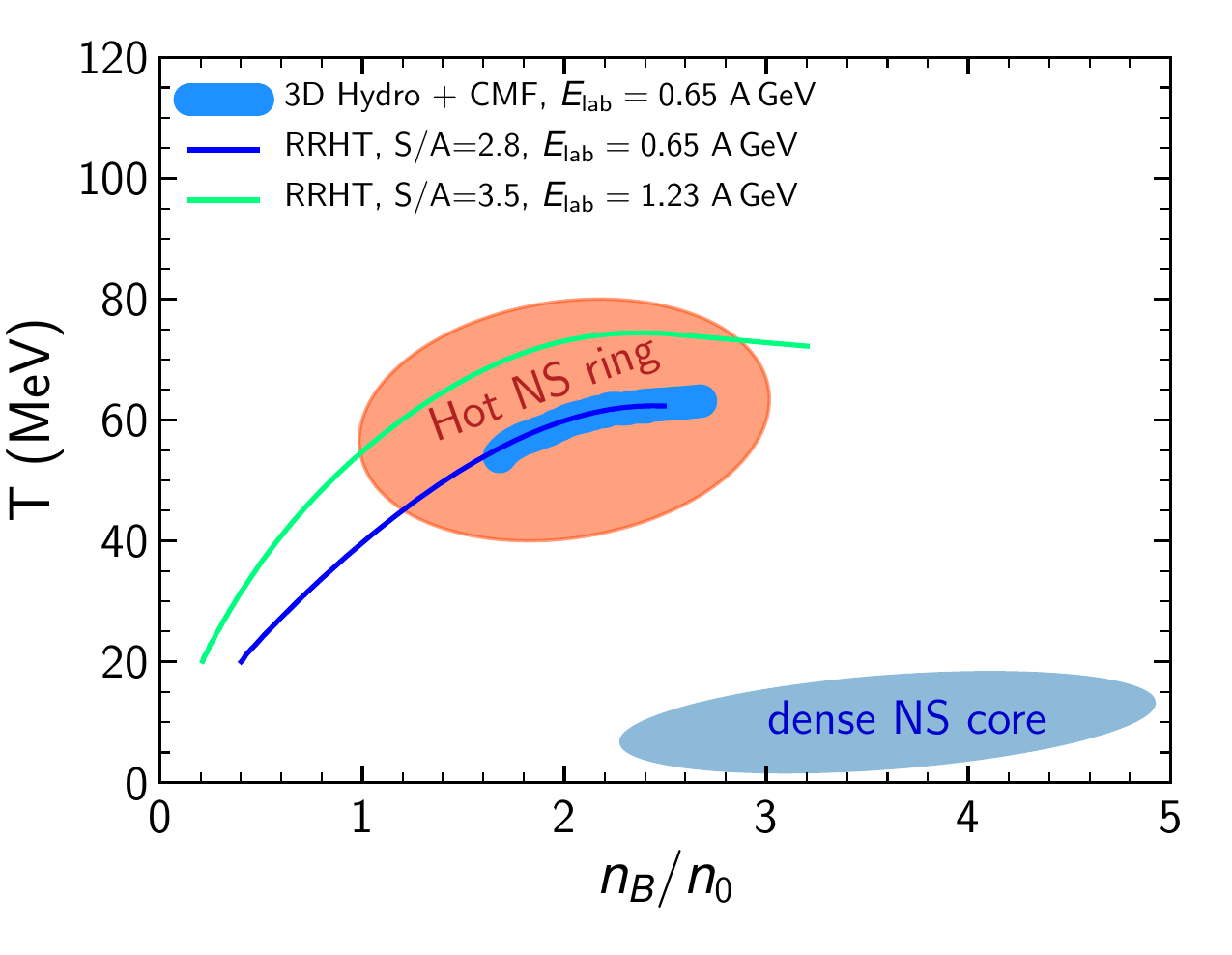}
\caption{Regions of $n_B-T$ phase diagram reachable in lower energy heavy ion collisions and in NS mergers. Bold line present results from full 3D hydrodynamic simulation with CMF EoS used. Thin lines indicate isentropes where entropy per baryon, $S/A$ was calculated by RRHT adiabat. Red and blue ellipses indicate regions of maximal temperature and density respectively, reachable in NS mergers, extracted from~\cite{Hanauske:2019qgs}.}
\label{fig:trajectories}
\end{figure}

\cref{fig:trajectories} shows the evolution of the baryon density $n_B$ and temperature $T$ in heavy ion collisions and regions accessible in NS mergers. The evolution of HICs is calculated in a detailed 3D hydrodynamic simulation as well as in a simplified 1D calculation with the RRHT adiabat. The regions reachable in NS mergers are obtained using general relativistic magneto-hydrodynamic simulations~\cite{Hanause:2018NS,Hanauske:2019qgs,Hanauske:2019new}. The figure illustrates similarities between hot and dense matter created in NS mergers and heavy ion collisions. The collisions have similarities due to rapid collision and compression of nuclear matter and differences due to spatial structure of NSs and presence of gravitational forces.

During the merger the neutron stars touch each other by their outer regions where densities are smaller than saturation density. At the initial stage of the merger these regions are rapidly compressed and heated, the hot NS regions appear shortly after $t\approx1$ ms. These regions are indicated by red in Fig.~\ref{fig:trajectories}, temperatures in these merged outer layers reach up to $T\approx75$ MeV. At the early stage of the merger there are two distinct dense regions that correspond to the NS cores. After the initial heating the differences between NS collisions and HIC emerge. The blue region in Fig.~\ref{fig:trajectories} is compressed much slower and is not heated so much. As the merger evolves further the two dense cold cores slowly move towards each other as a result of gravitation attraction. At the time around $t\approx5$ ms these two cold cores approach each other and form a dense and slightly heated, $T\lesssim20$ MeV, core of a newly formed rotating hypermassive NS. The hot region created at initial stage is split in two by the two approaching each other cores, so at $t\approx 5$ ms there are two hot spots and one dense cold core. During the further evolution these two hot spots rotate around the central core and at $t\gtrsim 15$ ms smear out into a hot NS ring. So at late postmerger stages the central region of hypermassive rotating NS is cold and dense and the middle interiors of NS consist of hot matter, blue and red regions of Fig.~\ref{fig:trajectories} respectively~\cite{Hanause:2018NS, Hanauske:2019qgs,Hanauske:2019new}.

\section{Summary}
\label{sec:summary}
A QCD equation of state consistent with lattice QCD data, heavy ion and neutron star physics is presented. The Chiral SU(3)-flavor parity-doublet Polyakov-loop quark-hadron mean-field model, CMF, is a phenomenological approach to the QCD thermodynamics with quark and hadron degrees of freedom. The QCD phase diagram calculated in CMF includes three critical regions: nuclear liquid vapor phase transition, phase transition relation related to chiral symmetry restoration, and a smooth transition to pure quark matter. In CMF quarks appear smoothly without discontinuities in the quark fraction. After chiral symmetry restoration takes place a significant quark fraction is present within the model, at chirally restored phase there is a mixture of quarks and hadrons. With the increase of temperature or density in the chirally restored phase, the quark fraction monotonously rises until the sudden appearance of quark matter. There the baryonic density is constituted only by free quarks, at this transition the thermodynamic quantities are continuous and baryon number susceptibilities experience non-monotonic behaviour.

The variety of QCD phenomena incorporated into the CMF model allows to reproduce LQCD data, neutron star astrophysical observables and apply the same EoS to heavy ion physics. This illustrates the similarity of underlying QCD physics in these physical phenomena and suggests that QCD matter can be probed experimentally not only in laboratory-based accelerator facilities, but also in stable neutron star observations and dynamical mergers of neutron stars. The later are observable by gravitational wave signals detectable at the LIGO and VIRGO laboratories.

\section*{Acknowledgements}
\label{sec:acknowledgements}
We acknowledge fruitful discussions with Matthias Hanauske, Elias Most, Jens Papenfort, and Luciano Rezzolla.
The authors thank HIC for FAIR, HGS-HIRe for FAIR, BMBF, and DFG for support.
J.S. appreciates the support of the SAMSON AG, WGG-Forderverein, and the C.W. F\"uck-Stiftungs Prize 2018. H.St. acknowledges the support through the Judah M. Eisenberg Laureatus Chair at Goethe University, and the Walter Greiner Gesellschaft, Frankfurt. Computational resources have been provided by the Center for Scientific Computing (CSC) at the J. W. Goethe-University, Frankfurt. 

\bibliography{cpod2018}

\end{document}